\documentclass[showpacs,aps,prl,twocolumn,superscriptaddress,floatfix]{revtex4-1}
\usepackage[english]{babel} 
\usepackage[T1]{fontenc} 
\usepackage[utf8]{inputenc} 
\usepackage{lmodern}
\usepackage{microtype}
\usepackage{hyperref}
\usepackage{amsfonts}
\usepackage{amsmath, amssymb,mathrsfs}
\usepackage{graphicx}
\usepackage{color}
\usepackage[normalem]{ulem}

\newcommand{\Gb}{{\langle \Gamma(t) \vert}}
\newcommand{\Pk}{{ \vert \Phi(t) \rangle}}
\newcommand{\cost}{c}
\newcommand{\xc}{{\bar x}}  

\newcommand{\llangle}{\langle \! \langle} 
\newcommand{\rrangle}{\rangle \! \rangle} 
\newcommand{\mm}{{\mu}} 
\newcommand{\smeq}{\!=\!}
\begin{document}

\title{An [imaginary time] Schr\"odinger approach to mean field games }

\author{Igor Swiecicki}
\affiliation{LPTMS, CNRS, Univ. Paris Sud, Université Paris-Saclay,
  91405 Orsay, France}  
\affiliation{LPTM, CNRS, Universit\'e   Cergy-Pontoise, 95302
  Cergy-Pontoise, France} 
\author{Thierry Gobron} 
\affiliation{LPTM, CNRS, Universit\'e   Cergy-Pontoise, 95302
  Cergy-Pontoise, France} 
\author{Denis Ullmo}
\affiliation{LPTMS, CNRS, Univ. Paris Sud, Université Paris-Saclay,
  91405 Orsay, France} 

\begin{abstract}
  Mean Field Games (MFG) provide a theoretical frame to model
  socio-economic systems.  In this letter, we study a particular class
  of MFG which shows strong analogies with the {\em non-linear
    Schr\"odinger and Gross-Pitaevskii equations} introduced in physics
  to describe a variety of physical phenomena.  Using this bridge many
  results and techniques developed along the years in the latter
  context can be transferred to the former, which provides both a new
  domain of application for the non-linear Schr\"odinger equation and
  a new and fruitful approach in the study of mean field games.  As an
  illustration, we analyze in some details an example in which the
  ``players'' in the mean field game are under a strong incentive to
  coordinate themselves.
\end{abstract}

\date{\today}
\maketitle

Mean field games, were introduced a decade ago by J-M.Lasry and
P-L. Lions \cite{LasryLions2006-1,LasryLions2006-2} and by M. Huang
and co-workers \cite{Huang2006} as a tractable version of game
  theory for a large number of players. This approach provides a very
versatile framework to model a vast range of socio-economic problems
ranging from social behavior
\cite{Dogbe2010,LachapelleWolfram2011,LaguzetTurinici2015,BesancDogguy2015}
to finance and economy
\cite{Lachapelle2015,GueantLasryLions2010,Achdou2014}.  Phrased in the
language of macroeconomy, it makes it possible to
go beyond the "representative agent" description
\cite{Achdou2014} and introduce, through its game-theory component,
some of the complexity associated with the variability of economic
agents' situations.  It does so while keeping some reasonable
  degree of simplicity thanks to the "mean-field" point of view
  taken.  In engineering science, it also proposes a manageable framework to approach complex optimization problems
involving a large number of coupled subsystems \cite{Huang2006}.

This relatively new field has witnessed a very rapid development in
the last few years, and has followed two major avenues.  The first one
is a mathematical approach in which one aims at proving the internal
consistency of the theory
\cite{CarmonaDelarue2013,BensoussanBook,Cardialaguet2015} as well as
deriving other rigorous results such as existence and uniqueness of
solutions for some classes of models
\cite{GomesSaude2014,Cardaliaguet-course}.  The other direction taken
was to develop efficient numerical schemes
\cite{Achdou2012,Gueant2011,LachapelleWolfram2011}. One thing which
has, however, prevented the diffusion of this tool at a significantly
larger scale is the lack of effective approximation schemes. In fact,
in spite of the ``mean-field-type'' assumptions, the constitutive
equations of these models remain rather difficult to analyze, in
particular because of their atypical forward-backward structure, and
only a few simple  models admit an analytical solution
\cite{Gueant2009,Bardi2012,LaguzetTurinici2015,Swiecicki2016}.  On the
other hand, full fledged numerical analyses of the mean field games
equations leave much to be understood.

We show here that there is a strong and deep relationship between mean
field games (or at least a large class of them), and the non-linear
Schr\"odinger (or Gross-Pitaevskii) equation, which has been studied
for almost a century by physicists to describe various physical systems ranging
from interacting bosons in the mean field approximation to gravity
waves in inviscid fluids.  The goal of this paper is to show that this
identification allows to transfer to mean field games (or at least to
a class of them) a vast array of knowledge and techniques that have
been developed through the years in this field (see
e.g. \cite{Yuri-RevModPhys1989,Kosevich1990,Kaup1990,PerezGarcia1997,Pitaevskii&StringariBook}).
In particular, this opens the way to very effective approximation
schemes leading both to a qualitative understanding and a good
quantitative description of the solutions of the mean field games
equations. This applies to many circumstances where a direct analysis
of the mean field games equations seems highly non-trivial, and in any
case has not been fully undertaken. As an illustration, we show how
this approach provides an essentially complete description of the
regime of strong, short range, attractive interactions, which is
presumably the most interesting case.

From a formal point a view, a mean field game is defined by two
components: the motion of the agents and the quantity they try to
optimize.  Each agent $i =1,\cdots N$ is assumed to be characterized
by a ``state variable'' $X_i(t) \in \mathbb{R}^n$, which, depending on
the problem under consideration, may represent physical space
\cite{LachapelleWolfram2011}, the amounts of some natural resources
\cite{GueantLasryLions2010}, or the position of a portfolio
\cite{Lachapelle2015}.  The dynamics of $X_i$ contains a deterministic
part which is controlled by the agent, and a random one associated
with external noise.  The simplest form of such a motion is a Langevin
dynamics
\begin{equation} \label{eq:Langevin}
d X_i = a_i(t) dt + \sigma dW_i \; ,
\end{equation}
where $W_i$ is a white noise of variance one.  On the other hand, each
agent chooses the drift $a_i(t)$ at time $t$ in order to minimize a
cost function whose typical form is: 
\begin{align} \label{eq:cost}
{\cost}[a_i]&(X_i(t),t) = \llangle {\cost}_T(X_i(T)) \rrangle_{\rm noise}  \\ 
 & + \llangle \int_{t}^T \left( \frac{\mm }{2 }a_i^2(\tau) -
  V[m_{\tau}](X_i(\tau)) \right) d\tau \rrangle_{\rm noise} \; . \nonumber
\end{align}
In this equation, $ \llangle \cdot \rrangle_{\rm noise}$ means an
average over the noise, $\mm >0 $ tunes the cost of a high drift
velocity, ${\cost}_T(x)$ is the final cost paid at the end of the
optimization period $T$, and $V[m_t](x)$ is both a function of $x$ and
a functional of the density of agents $m_t(x) \equiv \frac{1}{N}
\sum_i \delta (x - X_i(t))$.  Other forms of cost function or dynamics
can be introduced \cite{LasryLions2007,GomesSaude2014}; here, we shall
limit our discussion to the family of mean field games defined by
Eqs.~(\ref{eq:Langevin}-\ref{eq:cost}).

Defining the value function $u(x,t) \equiv \min_{a_i(.)}
{\cost}[a_i](x,t)$, the minimization of the cost function
Eq.~(\ref{eq:cost}), under the dynamics 
Eq.~(\ref{eq:Langevin}), leads to a system of coupled partial
differential equations \cite{LasryLions2006-2}: 
\begin{align} 
& \partial_t u  - \frac{1}{2\mm} {(\partial_xu)^2}  + 
\frac{\sigma^2}{2} \partial^2_{xx} u  = V[m_t](x) 
 \; , \label{eq:HJB} \\
&\partial_t m  + \partial_x (a(x,t) m)  -
\frac{\sigma^2}{2} \partial^2_{xx} m =0 \; , \label{eq:FP}
\end{align}
with $a(x,t ) \equiv - \frac{1}{\mu} \partial_x u(x,t)$.
Eq.~(\ref{eq:HJB}) is a Hamilton-Jacobi-Bellman (HJB) equation
propagating the value function $u(x,t)$ backward in time from the
final condition $u(x,T) \equiv {\cost}_T(x)$; Eq.~(\ref{eq:FP}) is a
Fokker-Planck (FP) equation propagating the density of agent $m_t(x) =
m(x,t)$ forward in time from the initial condition $m_0(x)$.  The two
equations (\ref{eq:HJB}) and (\ref{eq:FP}) are coupled due to the density
dependence of the "potential" $V[m_t](x)$ and by the fact that the
optimized drift $a(x,t)$ is the gradient of the value function.  

With a relatively simple change of variables \footnote{This variable
  change was actually already introduced in \cite{Gueant2011}.}, the
system, Eqs.~(\ref{eq:HJB}-\ref{eq:FP}), can be cast in a form which
we identify here as an {\em imaginary time} version of the {\em
  non-linear Schr\"odinger equation}.  As a consequence of this
  identification, we show hereafter that the associated formalism can
  be naturally introduced, leading to an effective approximation
  scheme. In particular, this approach relates to a very deep theorem
  derived by Cardialaguet and coworkers \cite{Cardaliaguet2013} which
  states that (under additional technical conditions) there exists an
  {\em ergodic state} $m^*(x)$ in the long time limit that the density
  $m(x,t)$ approaches for $T$ large when the time $t$ is sufficiently
  far from both $0$ and $T$.

To proceed, we introduce two new functions: $\Phi(x,t) = \exp [-
u(x,t)/{\mu \sigma^2}]$ (which corresponds to a Cole-Hopf transformation for
the HJB equation), and $\Gamma(x,t)  = m(x,t) / \Phi(x,t)$.
Eqs.~(\ref{eq:HJB}-\ref{eq:FP}) then read for these new variables:
\begin{align} 
- \mu \sigma^2 \partial_t \Phi  &= 
   \frac{\mu\sigma^4}{2} \partial^2_{xx} \Phi + V[m_t](x) \Phi
 \; ,  \label{eq:SchroPhi} \\
\mu\sigma^2 \partial_t \Gamma  &= 
   \frac{\mu\sigma^4}{2} \partial^2_{xx}\Gamma + V[m_t](x) \Gamma
 \; , \label{eq:SchroGamma} 
\end{align}
with the final condition $\Phi_T(x) \equiv \Phi(x,T) = \exp
[-u_T(x)/{\mu\sigma^2}]$ and the initial condition 
$\Gamma(x,0) \Phi(x,0) = m_0(x)$.

Under the formal replacement $\mu\sigma^2 \to -i \hbar$, these
equations are exactly those governing the evolution of a wavefunction
and its complex conjugate under the quantum Hamiltonian $\hat H = \hat
\Pi^2 / (2 \mm) + V[m_t](\hat X) $, where $ \hat \Pi \equiv
\mu\sigma^2 \partial_x$ and $\hat X$ are respectively momentum
and position operators.

For an arbitrary operator $\hat O = f(\hat X, \hat \Pi)$, let
us introduce the average
\begin{equation*} 
\langle \hat O \rangle(t)  \equiv \Gb  \hat O  \Pk = 
\int dx \, \Gamma(x,t) \hat O \Phi(x,t) \; , 
\end{equation*}
which, whenever $\hat O = O(\hat X)$, reduces to the classical mean
value $\int dx \, m(x,t) O(x)$. One has, as for the Schr\"odinger
equation, $\mu\sigma^2 \frac{d}{dt} \langle \hat O \rangle =
\langle [\hat H,\hat O] \rangle$. In particular, straightforward
algebra gives
\begin{equation}\label{eq:XPdot} 
\frac{d}{dt}  \langle \hat X \rangle  = 
     \frac{\langle \hat \Pi \rangle}{\mm} \; , \qquad
\frac{d}{dt} \langle \hat \Pi \rangle  = 
     \langle \hat F \rangle \; ,
\end{equation}
where we have introduced the ``force'' operator $\hat F[m_t]
  \equiv - \partial_x V[m_t](\hat X)$. The variance
  $\Sigma^2(t) \equiv \langle\hat X^2 \rangle - \langle\hat X
  \rangle^2$ evolves according to:
\begin{align}
\frac{d}{dt} & \Sigma^2   = \frac{1}{\mm} 
\left( \langle \hat X \hat \Pi + \hat \Pi \hat X \rangle  -  
  2\langle \hat \Pi\rangle \langle \hat X\rangle \right) 
\; , \label{eq:S2dot} \\
\frac{d^2}{dt^2} & \Sigma^2  =  \frac{2}{\mm^2} \left( \langle\hat\Pi^2\rangle -
\langle\hat\Pi\rangle^2 \right) \nonumber  \\
 & \quad - \frac{2}{\mm} \left( \langle \hat X \hat F)  \rangle 
-\langle \hat X \rangle  \langle  \hat F  \rangle
\right) \; . \label{eq:S2dotdot} 
\end{align}
If furthermore one considers potentials of the form
\cite{LasryLions2007} 
\begin{equation}
V[m_t](x) = U_0(x) + g \;m_t(x)^\alpha \; , \label{eq:PotInt}
\end{equation}
with $\alpha >0$, one gets explicitly
\begin{align}
\langle \hat F \rangle  & = \langle \hat F_0  \rangle  \equiv 
\langle -\nabla_x U_0(\hat X) \rangle \; \; ,
 \label{eq:F} \\
\langle \hat X \hat F  \rangle & 
= \langle \hat X F_0 \rangle  - 
\alpha  \langle H_{\rm int}  \rangle 
\label{eq:XF} \; , 
\end{align}
with $\langle H_{\rm int}  \rangle \equiv ({g}/{(\alpha+1)}) \int dx\;
m_t^{\alpha+1}(x)$; moreover the ``total energy''
\begin{equation} \label{eq:Etot}
\mathcal{E}(t)  \equiv \frac{1}{2\mm} \langle \hat \Pi^2 \rangle +
\langle U_0(\hat X)  \rangle +
\langle H_{\rm int}  \rangle 
\end{equation}
is a conserved quantity, i.e.\ $d \mathcal{E} /dt \equiv 0$.

Our claim is that Eqs.~(\ref{eq:XPdot}-\ref{eq:Etot}), together with
many results known in the context of the non-linear Schr\"odinger
equation, can form the basis of the analysis of a very large class of
mean field games for various associated potentials, including
some long range interactions.  In the following, we will illustrate
our point of view, restricting ourselves to the {\em one dimensional case} and
to potentials of the form Eq.~(\ref{eq:PotInt}) (though most of our
findings can be  extended straightforwardly to other cases).
We will furthermore focus mainly on the regime that we think is the
most interesting, namely the one of strong positive interactions ($g$
positive and large, in a sense clarified below).

To begin our analysis, it is presumably useful to start with
persistent solutions of
Eqs.~(\ref{eq:SchroPhi}-\ref{eq:SchroGamma}), which will eventually
correspond to the ``ergodic state''of Cardialaguet et al.\
\cite{Cardaliaguet2013}. These are obtained as $\Gamma(x,t)
= \psi^*(x) e^{\epsilon t/{\mu\sigma^2}}$ and $\Phi(x,t) =
\psi^*(x) e^{-\epsilon t/{\mu\sigma^2}}$, giving $m(x,t) =
m^*(x) = (\psi^*(x))^2$, where $\psi^*(x)$ is the solution of the time
independent non-linear equation $\hat H \psi^*(x) = \epsilon
\psi^*(x)$, that is
\begin{equation} \label{eq:GP}
\frac{\mu\sigma^4}{2 } \partial^2_{xx} \psi^* + U_0(x) \psi^* + g
(\psi^*)^{2\alpha +1} =  \epsilon \psi^* \; .
\end{equation}

We specialize from now on to $\alpha =1$ (the general case $\alpha
  > 0$ can be addressed following closely the approach described
  below \cite{SGU-2016}).  In this case Eq.~(\ref{eq:GP}) is exactly
  the (time-independent) Gross-Pitaevskii equation.  In the limit
  $U_0(x) = 0$ the lowest energy state is a soliton
  \cite{Pitaevskii&StringariBook}:
\begin{equation} \label{eq:WFsoliton}
\psi^*_{\rm s}(x)  =  {\frac{1}{\sqrt{2 \eta}}}
\cosh^{-1}\left( \frac{x}{ \eta} \right) \; ,
\end{equation}
with $\eta \equiv 2 \mu \sigma^4/g$, and
$\epsilon_{\rm s} =  {g}/{(4 \eta)}$.

Note that Eq.~(\ref{eq:WFsoliton}) provides a length scale, $\eta$, the
spatial extension of the soliton.  We now consider a non zero external
confining potential $U_0(x)$; by definition of a strong interaction
regime, the variations of $U_0(x)$ on a scale $\eta$ are small, that is
$|\eta \nabla_x U_0| \ll |\epsilon_{\rm s}|$ and $ |\eta\nabla_x^2
U_0| \ll |\nabla_x U_0|$.  Under these conditions, it is clear that,
away from $t=0$ and $t=T$ where the boundary conditions may force the
density of agents out of the soliton form, $m(x,t)$ will keep a
form close to $[\psi^*(x -\xc(t) )]^2$, centered around its mean
value $ \xc (t) \equiv \langle \hat X \rangle(t)$.  For this narrow
density profile one has $\langle \hat F_0 \rangle \simeq -
\nabla_x U_0(\xc )$, and applying Eq.~(\ref{eq:XPdot}) readily gives
\begin{equation} \label{eq:pdf}
\mu \frac{d^2}{dt^2} \xc(t)  = -\nabla_x U_0(\xc(t) ) \; .
\end{equation}
In the strong interaction regime, the motion of the soliton is simply
that of a classical particle of mass $\mu$ in the potential
$U_0(x)$.

The next point we need to address is the formation/destruction of the
soliton.  Indeed, considering for instance the neighborhood of $t \! =
\! 0$, the initial condition $m_0(x)$ can be taken far from the
soliton form, and one may ask how $m(x,t)$ evolves to it from
$m_0(x)$.  The short answer to this question is: ``quickly'' -- indeed
this process is dominated by interactions which are assumed to be
large.  To obtain further insight, let us assume that the density has
initially a Gaussian shape of variance $\Sigma^2_i$ and centered
around $\xc$. We use a Gaussian ansatz to describe its initial
evolution
\begin{equation*} 
m(x,t) \simeq  \frac{1}{\sqrt{2 \pi \Sigma^2(t)}} 
\exp\left[-\frac{(x-\xc)^2}{ 2 \Sigma^2(t)}\right] \; .
\end{equation*}
  Neglecting the influence
of the external potential during the formation of the soliton in
Eqs.~(\ref{eq:S2dotdot}-\ref{eq:XF}), and using that the total
energy Eq.~(\ref{eq:Etot}) is a conserved quantity, we can
express $\langle \hat \Pi^2\rangle /2\mu $ in terms of $\langle H_{\rm
  int} \rangle$ and its large $t$ stationary limit $\langle H_{\rm
  int} \rangle_*$ and obtain
\begin{align}
\frac{d^2}{dt^2} \Sigma^2   = & \;
\frac{2}{\mm} (\langle H_{\rm  int} \rangle_* - \langle H_{\rm  int}
\rangle)  \notag \\
 = & \; \frac{g}{2\mm\sqrt{\pi}}
  \left( \frac{1}{\Sigma_*} -\frac{1}{\Sigma(t)}
  \right) \; ,
\label{eq:dSigma2}
\end{align}
where $\Sigma_* = \sqrt{\pi} \eta$.   Imposing $\Sigma(t\!=\!0) =
\Sigma_i$, and introducing $z_t = \Sigma(t)/\Sigma_*$, $z_i =
\Sigma_i/\Sigma_*$, and
$\tau^* \equiv 2 \pi \sqrt{\mu \eta^3 / g}$,
Eq.~(\ref{eq:dSigma2}) can be integrated as
\begin{equation} \label{eq:FF}
- (z_t - z_i) - \log\left(\frac{1-z_t}{1-z_i} \right) =
\frac{t}{\tau^*}  \; .
\end{equation}

The destruction of the soliton can be tackled similarly, except that
the terminal condition imposed on $\Sigma^2$ is of the mixed form $\mu
\mathrm{d} \Sigma^2 / \mathrm{d}t (T) + 2 (\partial^2_{xx} \cost_T(\xc
(T))) \Sigma^2(T) = \sigma^2$, and thus gives a  different
expression (not shown) for the solution of Eq.~(\ref{eq:dSigma2}).  
 One finds that $\Sigma(T) \simeq
\Sigma^{*}(1+\xi)$ where $\xi \simeq 0.43$ is a number. So the final
density  $m(x,T)$ has a dispersion which remains of order $\Sigma_*$.

Setting aside the precise way the soliton is formed or destroyed near
the boundaries $t \!=\!0$ and $t \!=\!T$, the important point here is
that the characteristic time $\tau^* = \pi \eta
  \sqrt{\mu/|\epsilon_{\rm s}|}$ which emerges is short, in the sense
that $\eta$ is assumed the smallest length scale of the problem and
$\epsilon_{\rm s}$ the largest energy scale of the problem.  This is
consistent with the fact that during its formation, the soliton can be
considered immobile and centered around $\xc$.  The terminal condition
on the other hand does not involve directly $m(x)$ as what is fixed is
the final cost function $\cost_T(x)$.  Using again
that near $T$ the density remains localized on a scale $\sim \Sigma_*
\sim \eta$ which is short, one can show however that one has for the
center of the soliton $\xc (t)$ the terminal condition
\begin{equation} \label{eq:terminal} 
\mu \frac{d\xc}{dt}(T) = - (\partial_x \cost_T)[\xc (T)]
\; .
\end{equation}


%

\begin{figure}[ht]
\includegraphics[width=7.7cm]{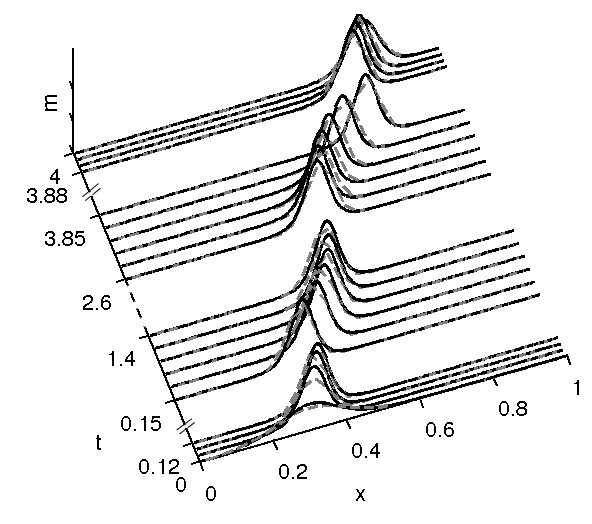}
\caption{Solution of the mean field game equations for the density of
  agents in a typical configuration.  Solid black: numerical solution
  of Eqs.~(\ref{eq:HJB}-\ref{eq:FP}); Dashed grey : Solution for the
  Gaussian ansatz solution of Eqs.~(\ref{eq:pdf}-\ref{eq:FF}). The
  initial density is $m(x,t \smeq 0)= \frac{1}{2 \eta_i}
  \cosh^{-2}\left( \frac{x-x_i}{ \eta_i} \right) $ with $x_i=0.3$,
  $\eta_i=0.2$, and the final cost $c_T(x)= 2 \pi^2(x-0.8)^2$.  The
  parameters of the model are $\sigma=0.45$, $\mu=1$, $T=4$; the
  potential is as in Eq.~(10) with $\alpha=1$, $g=2$ and $U_0(x) =
    - (\pi^2/8) (x - 0.5)^2$. To make more visible the creation and
  relaxation of the soliton the time scale of the initial and final
  time periods have been magnified.}
\label{fig:comparison}
\end{figure}
As an illustration, we show in Fig.~(\ref{fig:comparison}) a
comparison, for a rather typical setting, between a numerical solution
of Eqs.~(\ref{eq:HJB}-\ref{eq:FP}) for a potential as in
Eq.~\eqref{eq:PotInt} with $\alpha=1$ and the  predictions
derived from the above analysis.  The quantitative agreement is seen
to be very good.  More generally, we can now give a fairly complete
description of the solution of the mean field game equations in the
regime of strong short-ranged positive interactions that we consider
here.  One can distinguish three distinct periods of time.

In the first one (the ``formation of the soliton''), the agents
coordinate themselves through their strong mutual interaction and
evolve from  an arbitrary initial distribution
$m_0(x)$ to a localized one whose extension $\Sigma_*$ results from a
balance between the agents' interaction (which tends to reduce $\Sigma_*$), and noise (which tends to increase it).  In this phase, which takes
place on the shortest time scale $\tau^*$, the external potential
$U_0(x)$ plays little role, and the final utility $\cost_T(x)$ no role
at all.  Whenever the Gaussian ansatz is accurate during this phase,
Eqs.~(\ref{eq:dSigma2}-\ref{eq:FF}) provide a quantitative
description of the time evolution of the density of agent.  If
  $m_0(x)$ is not well approximated by a Gaussian this description is
  presumably a bit more qualitative.  Note however that the only
place where the Gaussian form has been explicitly used here is when
expressing $\langle H_{\rm int} \rangle$ in terms of the variance
$\Sigma^2(t)$ of $m(x,t)$.  As long as this relation is approximately
maintained, and given that $m(x,t)$ has to converge to the soliton
form which is well approximated by a Gaussian, the description
Eqs.~(\ref{eq:dSigma2}-\ref{eq:FF})) should be reasonably accurate.

The third (and last) time period extends also over the short time
scale $\tau^*$ just before $T$, when the agents density slightly relax
from the soliton form to adjust to the final cost function
$\cost_T(x)$.  Since the boundary condition does not involve the final
density $m(x,T)$  one can assume there a compact form for $m(x,t)$ with a
finite spread on a scale $\sim \Sigma_*$.  During this phase, the
external potential $U_0(x)$ plays little role, and the initial
  density of agents no role at all.

In between, assuming of course $T \gg \tau^*$, most of the time period
$[0,T]$ is characterized by the relatively slow motion of the agents
following Eq.~(\ref{eq:pdf}).  Because $\tau^*$ is so short, and
because the dynamics of $\langle \hat X \rangle$ and $\langle \hat \Pi
\rangle$ are controlled by the external potential $U_0(x)$, their
values barely move during the formation or the destruction of the
soliton, and thus Eq.~(\ref{eq:pdf}) can be assumed to be valid all along
$[0,T]$. Therefore the details of the dynamics in the initial and
final phase of the formation of the soliton will not change
drastically what will happen during the soliton propagation.

In the intermediate phase, the dynamics is therefore determined: by
$m_0(x)$, which fixes the initial position of the soliton; by
$\cost_T(x)$, which sets the final velocity of the soliton; and
by the confining potential $U_0(x)$ which drives the motion between
the two.  We arrive thus at this relatively non-intuitive result that 
{\em the details of the strong coordination between the agents,
  which is assumed to be the largest force at work here, plays little
  role in the global picture.}

Considering now the long time limit studied by Cardialaguet and
coworker \cite{Cardaliaguet2013}, the picture we obtain is the
following: the simplest way to form a trajectory fulfilling the
boundary condition $\xc = \xc_0$ at $t \smeq 0$ and
Eq.~(\ref{eq:terminal}) at $t \smeq T$ for very large $T$, is to use an
initial velocity $\dot\xc_0$ such that the energy $E \equiv \mu
\dot \xc_0^2/2 + U(\xc_0)$ is almost equal to $U_0(x_{\rm max})$, with
$x_{\rm max}$ the maxima of $U_0(x)$ (which is thus an unstable fixed
point).  In this way, the trajectory reaches $x_{\rm max}$ with an
 almost zero velocity, thus staying there for an arbitrarily
long time, before picking speed again to fulfill
Eq.~(\ref{eq:terminal}) at $t \smeq T$.  The ergodic state appears in
this way as $m_*(x) \equiv [\psi^*(x -x_{\rm max})]^2$, and is approached
exponentially quickly if $U_0(x )$ is at least quadratic around $x_{\rm ma
x}$.   

We stress however that dealing with a boundary condition problem
(implying initial and final times) rather than an initial value
problem (initial position and velocity fixed) considerably changes
things compared to classical mechanics, especially with respect to the
uniqueness of the solution.  Indeed, if there is more than one local
maxima of $U_0(x)$, one can in most circumstances build more than one
solution to the problem (depending on the energy $U_0(\xc (t=0))$ and
$U_0(\xc (t=T))$, and on the location of the local maxima relative to
$\xc (t=0)$ and $\xc (t=T)$).  Taking the solution associated with the
lowest value of the cost function Eq.~(\ref{eq:cost}) will make it
possible to select the correct one, but this process should imply some
phase transition as $T$ increases as the system shifts from one local
maxima to a higher one.

In this letter, we have stressed a natural connection between
non-linear Schr\"odinger equations and mean field games expressed by
Eqs.~(\ref{eq:SchroPhi}-\ref{eq:SchroGamma}) which
makes possible the transfer to this latter field of a large variety of
tools to analyze, both qualitatively and quantitatively, a wide class
of systems which appear significantly more difficult to address
directly in the original form.
We have focused on the regime of strong short-ranged
interactions but other cases (long range interactions, strong
confining potential), and higher dimensional problems, could be
addressed very similarly.  Exploiting fully this connection
  provides both a new playground for physicists familiar with the
  non-linear Schr\"odinger equation and a path to powerful
  approximation schemes for mean field games equations.  
    The analysis of real socio-economic problems should eventually
  benefit from these progresses.

{\em Acknowledgments:} This research has been conducted as part of the
project Labex MME-DII (ANR11-LBX-0023-01).  We thank Nicolas Pavloff
for introducing us to the subtleties of the non-linear Schr\"odinger
equation and Steven Tomsovic for helpful discussions.

%

\end{document}